# QGRP: A Novel QoS-Geographic Routing Protocol for Multimedia Wireless Sensor Networks.

Mohammed-Amine KOULALI[1], Mohammed EL KOUTBI[1], Abdellatif KOBBANE[1]
and Mostafa AZIZI[2].

[1] Laboratoire des Systèmes d'Information Mobiles (SIME), Mohammed V-Souissi University, ENSIAS
Rabat, Madinat Al Irfane, BP 713, Agdal, Rabat, Mororcco

[2] Laboratoire MATSI, Université Mohammed I, ESTO
Oujda, 60000, Morocco

**Abstract**
Thanks to the potential they hold and the variety of their application domains, Multimedia Wireless Sensor Networks (MWSN) are forecast to become highly integrated into our daily activities. Due to the carried content nature, mainly composed of images and/or video streams with high throughput and delay constraints, Quality of Service in the context of MWSN is a crucial issue. In this paper, we propose a QoS and energy aware geographic routing protocol for MWSN: QGRP. The proposed protocol addresses bandwidth, delay and energy constraints associated with MWSN. QGRP adopts an analytical model of IEEE 802.11 Distributed Coordination Function (DCF) to estimate available bandwidth and generates loop-free routing paths.

## 1. Introduction

Multimedia Wireless Sensor Networks [1-2] are a class of Wireless Sensor Networks where the carried flows are mainly composed of images and/or video streams. With the availability of low cost CMOS cameras and microphones, MWSN are forecast to become highly integrated into our daily activities [3] with applications spanning over a large panel of domains and contexts: from healthcare and intelligent patient monitoring to disaster relief and industrial process supervision through intrusion detection and military deployments, MWSN hold a promising future.

Being mainly data centered networks, efficient routing of data packets from sensors to base station is the key stone of any viable MWSN deployment. Thus, optimal routing is crucial during both query and data dissemination phases [4] the carried multimedia content is subject to strict QoS requirements. Consequently, QoS routing protocols for MWSN should account for their specificities and depend on their application context.

As depicted in Fig.1, MWSN are built by aggregating hundreds to thousands of energy constrained/battery powered devices. Since battery replacement is not always an option (deployment in hostile areas such as battle fields or chemical facilities) and energy harvesting techniques have not reached the maturity level allowing their deployment at commercial scale: energy consumption should be kept minimal by reducing communication and computing operations. Also, to reduce production costs, storage and computing capabilities are limited. Thus, collaborative distributed computing is a necessity for MWSN to operate optimally. Also the bandwidth scarcity and variable channel quality associated with multimedia content are key characteristics of MWSN.

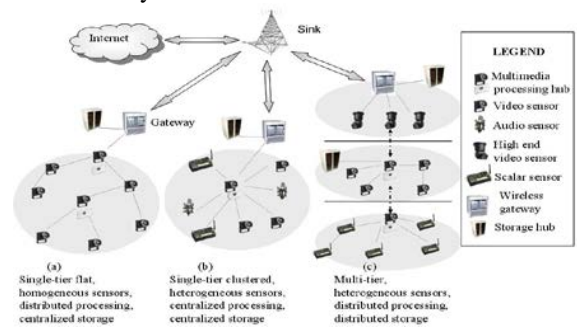

Fig. 1 Hierarchical Wireless Multimedia Network [1].

We focus on the bandwidth and delay constraints which happen to be critical issues because of the amount and nature of carried multimedia content. The energy scarceness is accounted for during route establishment and the routing overhead is kept minimal to reduce energy consumption. As Location awareness could be of great help when routing decisions are taken. We propose a composite routing metric that accounts for geographic progress along with previously cited criteria.

In this paper we propose a protocol that addresses the issue of QoS routing in wireless sensor networks subject to bandwidth and delay constraints and accounts for energy consumption. To make more accurate estimation of the available bandwidth, we adopt an analytical model of IEEE 802.11 DCF. Also, in order to guarantee loop-free routing paths we introduce destination sequence numbers and finally energetic performances of QGRP are provided in terms of average residual energy, energy efficiency and standard energy deviation.

The paper is organized as follows: Section 2 presents related works. Our proposed protocol is described in section 3. Some simulation results are presented and discussed in section 4. Finally we conclude the paper and announce our future works in Section 5.

## 2. PREVIOUS WORK

Depending on when routes are constructed, routing protocols are classified into reactive, proactive and hybrid. Whereas proactive routing protocols such as SAR [7] periodically construct and maintain routing tables increasing consequently the communication and storage overhead. Reactive routing protocols notably: LABQ [5] and MMSPEED [6] construct routes on demand. Hybrid protocols such as ZRP [8] combine both reactive and proactive paradigms.

The aspects of localization and QoS [9] seem to be judicious to use for routing [10] in MWSN. While geographic routing protocols focus on delay optimization by greedily forwarding packets towards sensors near the destination increasing consequently their progress. QoS routing protocols [11-12] on the other hand concentrate on optimally exploiting the scarce resources (energy, computing capabilities) in order to achieve packets routing while respecting QoS constraints (delay, delay-jitter, bandwidth). The main difference between geographic protocols is their definition of progress. MFR [13] and compass routing [14] are representatives of this category.

Some propositions have been made to combine geographic and QoS routing notably SPEED [15], MMSPEED [6] and [16]. These protocols are mainly concerned with the delay constraint. SPEED does not support traffic differentiation. MMSPEED does not account directly for energy as a routing metric and adopts probabilistic forwarding to achieve load balance. The authors of [17] propose the use of a combined metric accounting for progress along with other QoS factors: energy and intra-sensor packets sojourn time.

## 3. PROPOSED PROTOCOL

In order to enhance the accuracy of available bandwidth estimation we adopted the analytical model of IEEE 802.11 DCF proposed in [23]. Thus, we derived analytically the conditional collision probability and average back-off duration. Table 1 presents symbols used in proposed protocol description.

Table 1: List of symbols used in proposed protocol description

| Symbol | Description |
|---|---|
| $P_a$ | Transmission attempt probability. |
| $P_c$ | Conditional collision probability. |
| $CW_{max}$ | Contention window size. |
| $CW_{min}$ | Minimal contention window size. |
| $IN_r$ | Region silenced when sensor r transmits. |
| $CS_s$ | Carrier Sense region of sensor s. |
| $B_{U,V}$ | Available bandwidth on link (U,V). |
| $E_V$ | Residual energy of sensor V. |
| $E_{Vin}$ | Initial energy of sensor V. |
| $B_{no}$ | Nominal channel capacity. |
| $T_v$ | Duration of a virtual slot. |
| $V$ | Duration of the protocol header and payload. |

### 3.1 Available bandwidth estimation

Available bandwidth estimation techniques fall into active and passive categories. While the first set of techniques [18-19] relies on injecting end-to-end probe packets in the network to evaluate parameters such as transmission delay for various emission rates. The latter [20-21] exploits information such as: sender/receiver idle periods synchronization, back-off periods, collisions probability etc. The passive approach does not modify the network status and consequently produces no overestimation of the available bandwidth.

We enhance the approach of [20] for passive available bandwidth estimation through the use of analytically inferred $P_c$ and average back-off duration values instead of the trace file estimated ones. Bianchi presented an analytical model of IEEE 802.11 DCF in [22] for single cell networks. The model was generalized in [23] to multi-hop networks. [22] and [23] assume saturation condition : The terminal has always a packet to transmit. In our case, sensors exchange periodically hello packets and they either transmit or forward control/data packets. Thus, we consider the saturation assumption satisfied.

We derive the conditional collision probability from [23] by solving 2 non linear fixed point equations system:

$$P_a = \frac{2 - 4P_c}{(1 - 2P_c)(CW_{max} + 1) + P_c CW_{min}(1 - (2P_c)^m)} \quad (1)$$

$$P_c = 1 - (1 - P_a)^{|CS \cap IN_r|}(1 - P_a)^{|CS - IN_r|\frac{V}{T_v}} \quad (2)$$

With $m = \log(\frac{CW_{max}}{CW_{min}})$, we also assume that the region silenced by the receiver $r$ : $IN_r$ is covered by the sender carrier sense region $CS_s$, thus the conditional collision probability reduces to:

$$P_c = 1 - (1 - P_a)^{|CS \cap IN_r|} \quad (3)$$

Solving fixed point equations is both energy and computationally expensive. Instead of doing on-line computation of the conditional collision probability, we solve the system off-line for various sender/receiver distances and network densities. Table 2 summarizes the obtained results.

Table 2: $P_c$ for various density/distance configurations

| Density | Distance | | | |
|---------|----------|------|------|------|
|         | 100m     | 150m | 200m | 200m |
| 90      | 0.1444   | 0.2535 | 0.3319 | 0.3910 |
| 100     | 0.1781   | 0.2727 | 0.3436 | 0.4062 |
| 110     | 0.1781   | 0.2727 | 0.3544 | 0.4198 |
| 120     | 0.1781   | 0.2898 | 0.3739 | 0.4323 |

Each sensor infers its conditional collision probability through weighted average of $P_c$ values of the two nearest configurations stored in its memory. The weights indicate how much the actual sender/receiver configuration diverges from the stored ones in terms of distance.

3.2 Composite routing metric

For a sensor U, the set of potential forwarders $fw_U$ is composed by neighbors respecting the bandwidth constraints that form with U and the base station an angle lesser or equal than $\pm\frac{\pi}{2}$.

To account for geographic progress, available bandwidth and energy consumption while establishing a routing path to the base station, we use a composite link routing metric. The metric value for the link (U,V) with $V \in fw_U$ is given by:

$$P_c = \frac{1}{r * \theta}(\alpha \frac{B_{U,V}}{B_{no}} + \beta \frac{E_V}{E_{V_{in}}}) \quad (4)$$

$\alpha$ and $\beta$ are weighting factors related to each other by the formula : $\alpha + \beta = 1$. $r$ is the Euclidean distance separating sensor V from the base station. $\frac{B_{U,V}}{B_{no}}$ and $\frac{E_V}{E_{V_{in}}}$ are respectively available bandwidth and residual energy ratios.

Maximum geographic progress is offered by neighbors closer to the base station that induct less deviation from the virtual straight line linking the sensor performing routing decision to the base station.

3.3 Route establishment & admission control

Destination sequence numbers are used to obtain loop-free routing paths to the base station. A route is said to be fresher than another if it has greater destination sequence number or has the same sequence number but with higher path bandwidth. In order to obtain a path to the base station the sensors proceed as follows:

The source unicasts a Route Request (RREQ) packet to its neighbor insuring the highest value of the composite link metric. The RREQ packet contains the flow required bandwidth and the available bandwidth on the link used to forward it.

- When a RREQ packet is received, the presence of a fresh routing entry to the destination is verified. If the receiver happens to be the base station or has a fresh routing entry to it, then the receiver replies with a Route Reply (RREP) packet
- If no routing entry is found the RREQ is forwarded as described in the first step.
- RREQ (respectively RREP) packets record the partial path bandwidth on their journey to the base station (respectively source). Thus, intermediate sensors update their routing entry information for the base station (respectively source).
- RREQs that are not followed by RREPs for a fixed period are remitted till a maximum limit is reached.

Admission control [24] is realized in order to preserve previously admitted flows from performance degradation. When a sensor receives a route request that cannot be fulfilled due to due to bandwidth requirement violation, it sends back a notification packet to the source containing the maximal bandwidth that can be granted. Upon receiving the notification packet, the source could either reduce its requirements or schedule a future route request hoping that the available bandwidth will increase. If the source chooses to wait for more available bandwidth, data packets will be buffered.

## 4. SIMULATIONS

We simulate the proposed protocol on the NS 2.34 network simulator [25] along with AODV [26]. Although AODV is an AdHoc routing protocol, many articles use it to benchmark proposed protocols for sensor networks. Furthermore, several works try to adapt AdHoc routing protocols to wireless sensor networks. AODV is one of the simplest and lightweight AdHoc routing protocols, it seems to be a good candidate for this operation of adaptation. So, AODV will be used in this paper to evaluate our proposed approach. Our ideas can be eventually embedded in other existing routing protocols. We benchmark QGRP against AODV on the basis of six averaged performance metrics: Throughput, Packet Delivery Ratio (PDR), End-to-End Delay, sensors Residual Energy, Energy Efficiency and Standard Energy Deviation. The latter two parameters are defined as follows:

- Energy Efficiency: The ratio of total amount of energy dissipated by all source and forwarder sensors to the number of unique packets received by base station.
- Standard Energy Deviation: The average variance between the residual energy levels on all sensors.

For the simulations, nominal bandwidth is set to 2 Mbit/s and transmission range fixed to 250 m. The simulations were conducted for topologies of 90, 100, 110 and 120 sensors, with identical initial energy of 40 J, randomly and uniformly scattered in a $10^6$ m$^2$. Three flows, respectively, 0.5, 0.4 and 0.2 Mbit/s were injected in the network. The simulation was repeated 10 times for each topology and each time the source sensors and the base station changed. The obtained results were averaged to reduce the chances that the observations are dominated by a certain scenario which favors one protocol over another. $\alpha$ and $\beta$ were set experimentally and their values correspond to the ones that ensure optimal performances of QGRP : $\alpha = 0.7$, $\beta = 0.3$.

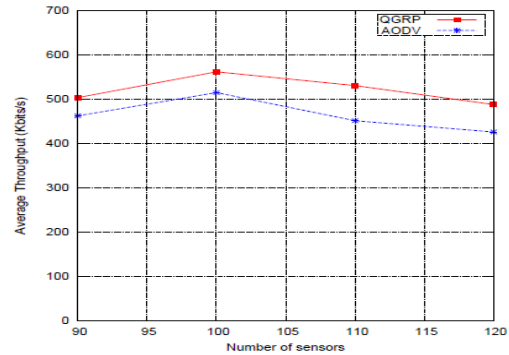

Fig. 2 Average Throughput, QGRP vs AODV.

Fig.2 shows that the proposed protocol outperforms AODV in terms of achieved Throughput. The proposed protocol manages to enhance the throughput by 17,56% for the topology with 110 nodes. The proposed protocol achieves a maximal throughput of 530,155 Kbit/s with a network density of 100 sensors. AODV realizes its worst performance with a topology of 120 nodes and manages to achieve only 425,503 Kbit/s. While, in QGRP, sensors violating the bandwidth constraints are pruned and neighbors with maximal bandwidth adopted as forwarders. AODV selects the path with the shortest hop count, regardless of the available bandwidth that its nodes can grant to admitted flows. Also, admission control performed by QGRP prevents degradation of previously admitted flows.

We notice that even with a nominal channel capacity of 2 Mbit/s, collisions, congestion and intra-flow contention reduce significantly the network average throughput.

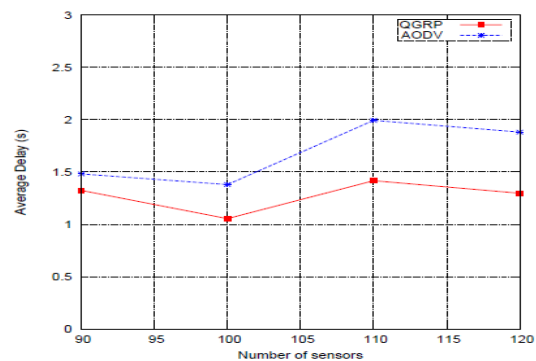

Fig. 3 Average Delay QGRP vs AODV.

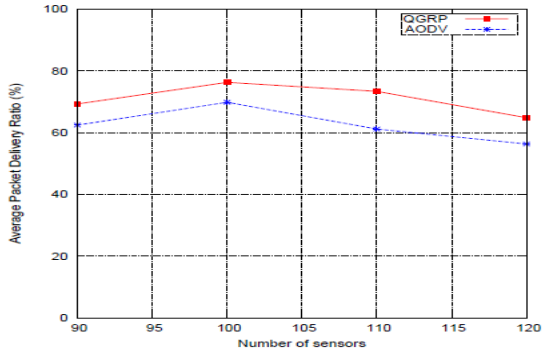

Fig. 4 Average PDR, QGRP vs AODV.

The composite routing metric is maximized by choosing neighbors closer to the virtual line linking the sensor to the base station. This reduces the Euclidean distance that packets have to travel in order to reach the base station and consequently shortens the associated transmission delay. As shown in Fig.3 the proposed protocol manages to reach a delay of 1.05s for the topology with 100 sensors whereas AODV reaches only 1.38s. For the topologies with 110 and 120 sensors AODV is outperformed by respectively 45% and 40,65%. Also, the PDR results depicted in Fig.4 correlate with average throughput results. QGRP attains a PDR of 76,35% for the topology of 100 sensors.

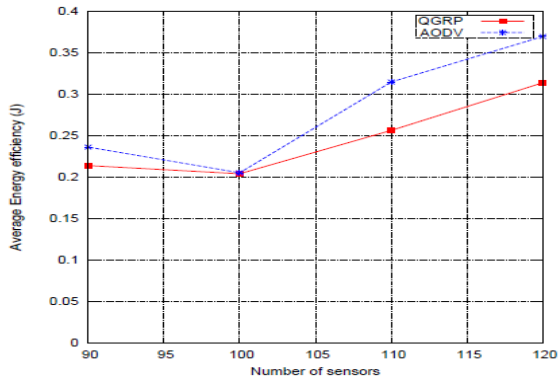

Fig. 5 Average Energy Efficiency, QGRP vs AODV.

The proposed protocol accounts for sensors residual energy while performing routing decisions. Thus, sensors having consumed less energy will be privileged. Also, the load will be shared among sensors to avoid the rapid depletion of certain sensors batteries and the disconnection of the network. Fig.5 shows that QGRP is more energy efficient than AODV. For topologies with 110 and 120 sensors the enhancements are respectively: 22,76% and 17,74%. More data packets reach the base station with less energy consumption.

Fig.6 illustrates the average residual energy of the network sensors at the end of the simulation, whereas Fig.7 depicts the standard energy deviation from this average. The proposed protocol consumes less energy than AODV for the topologies 90, 110 and 120 even though it ensures higher throughput, PDR and smaller transmission delay to the carried flows. With regard to the topology of 100 sensors, results realized by AODV are justified by the fact that QGRP realizes its best performances in that topology which results in much more energy consumption. Poor AODV realized results for this topology allows it to handle carefully its energetic resources. QGRP manages to balance the load over the network sensors resulting in smaller or equal Average standard energy deviation than AODV with net advantage in terms of other performance metrics.

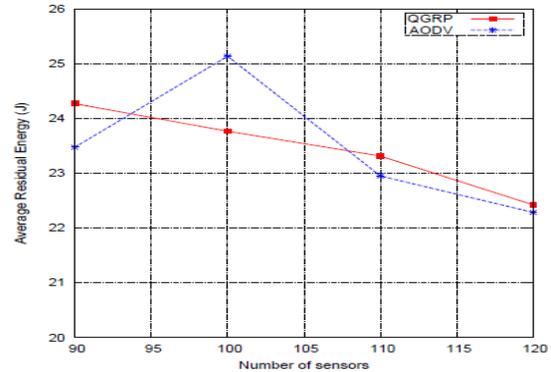

Fig. 6 Average Residual Energy, QGRP vs AODV.

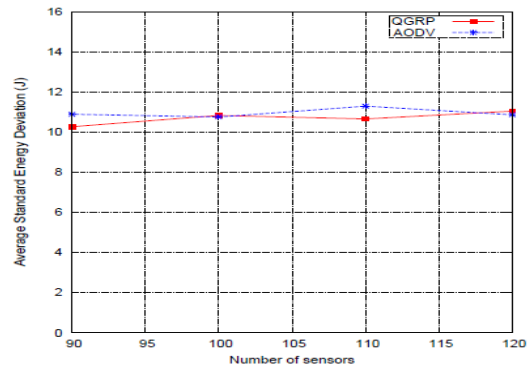

Fig. 7 Average Standard Energy Deviation, QGRP vs AODV.

## 5. CONCLUSION

In this paper we've proposed QGRP: a QoS Geographic routing protocol for Multimedia Wireless

Sensor Networks. We've conducted simulations to benchmark the proposed protocol against AODV on the basis of: average delay, throughput, PDR, energy efficiency, standard energy deviation and average sensors residual energy parameters. The results establish the QGRP performances. We plan to enhance the proposed protocol by means of:

- Hybridization through the development of a proactive component.
- Dynamic adjustment at sensor level of $\alpha$ and $\beta$ factors on the basis of observed network status.
- Traffic differentiation for scalar and multimedia content since bandwidth is less critical for the first category.
- Support of heterogeneous battery capacities.

**Mohammed-Amine KOULALI** graduated from ENSAO (National Higher School for Applied Science, Oujda) in 2006, holder of a Master degree in advanced computing and applications from the Franche comté University of Besançon, France (2009). He is currently a "Ph.D. candidate" at ENSIAS Under the supervision of Prof. Dr. Mohammed EL KOUTBI. His research activities focuse on QoS routing in wireless sensor networks, trust networks and mathematical frameworks for modeling and analyzing Ad-hoc and sensor networks. He served as a reviewer for international conferences such ICC and Globecom. He teaches computer science at ENSAO since 2010.

**Dr. Mohammed El Koutbi** is presently working as a full professor at the University Mohamed V -Souissi at ENSIAS (an engineering school in computer science). He held a PhD degree in Computer Science from Montreal University, Canada since July 2000. He has been heavily involved in teaching, research, curriculum development, certification activities, lab developments, and faculty/ staff hiring at the department. He has several years of strong industrial and research experience in software engineering and computer networking. His current research is in the area of computer networking (mobile computing and Adhoc networks) and software engineering.

**Dr. Abdellatif Kobbbane** is currently an Associate Professor (Professeur Assistant) at the communication networks department of Ecole Nationale Suprieure d'Informatique et d'Analyse des Systemes (ENSIAS), Mohammed V-Souissi University, Morocco since 2009. He received his PhD degree in computer science from the Mohammed V-Agdal University (Morocco) and the University of Avignon (France) in September 2008. He received his research MS degree in computer science, Telecommunication and Multimedia from the Mohammed V-Agdal University (Morocco) in 2003. His research interests lie with the field of wireless networking, performance evaluation in wireless network and NGN (DTNs, Mesh networks, cognitive radio, Mobile Social networks, ...). Doctor Abdellatif Kobbane is a member of SIME Lab in MIS research group leader. He served as a r eviewer for many international journals and conferences such COMCOM, ICC, Globecom, WiMob, WCNC, and l oc. Dr Kobbane is an I EEE member and President of the Association of Researchers in Information Technology in Rabat Morocco.

**Dr. Mostafa Azizi** received his diploma of State engineer in Automation and Industrial Computing in 1993 f rom the Mohammadia School of engineers at Rabat (EMI) and obt ained his PH.D. in Computer Science in 2001 from the University of Montréal (DIRO-FAS-UdeM). He is currently professor of computer science at the University of Oujda. His research interests include: Verification/Coverification of real-time and embedded systems, Data communication and s ecurity, and Computer-aided management of industrial processes.